\input harvmac
\newcount\figno
\figno=0
\def\fig#1#2#3{
\par\begingroup\parindent=0pt\leftskip=1cm\rightskip=1cm\parindent=0pt
\global\advance\figno by 1
\midinsert
\epsfxsize=#3
\centerline{\epsfbox{#2}}
\vskip 12pt
{\bf Fig. \the\figno:} #1\par
\endinsert\endgroup\par
}
\def\figlabel#1{\xdef#1{\the\figno}}
\def\encadremath#1{\vbox{\hrule\hbox{\vrule\kern8pt\vbox{\kern8pt
\hbox{$\displaystyle #1$}\kern8pt}
\kern8pt\vrule}\hrule}}

\overfullrule=0pt

%macros
%
\def\tilde{\widetilde}
\def\bar{\overline}
\def\Z{{\bf Z}}

\def\S{{\bf S}}
\def\R{{\bf R}}

\font\zfont = cmss10 %scaled \magstep1
\font\litfont = cmr6

\def\bigone{\hbox{1\kern -.23em {\rm l}}}
\def\ZZ{\hbox{\zfont Z\kern-.4emZ}}
\def\half{{\litfont {1 \over 2}}}

\Title{hep-th/9803131, IASSNS-HEP-98/21}
{\vbox{\centerline{Anti-de Sitter Space, Thermal Phase}
\bigskip
\centerline{Transition, And Confinement In Gauge Theories}}}
\smallskip
\centerline{Edward Witten}
\smallskip
\centerline{\it School of Natural Sciences, Institute for Advanced Study}
\centerline{\it Olden Lane, Princeton, NJ 08540, USA}\bigskip

\medskip

\noindent

The correspondence between supergravity (and string theory)
on $AdS$ space and boundary conformal field theory relates
the thermodynamics of ${\cal N}=4$ super Yang-Mills theory in four
dimensions to the thermodynamics of Schwarzschild black holes in
Anti-de Sitter space.  In this description, quantum phenomena
such as the spontaneous breaking of the center of the gauge
group, magnetic confinement, and the mass gap are coded in classical
geometry.  The correspondence makes it manifest that the entropy
of a very large $AdS$ Schwarzschild black hole must scale
``holographically'' with the volume of its horizon.  By similar
methods, one can also make a speculative proposal for the description
of large $N$ gauge theories in four dimensions without supersymmetry.

\Date{March, 1998}
%text of paper

\def\N{{\cal N}}
\newsec{Introduction}

Understanding the large $N$ behavior of gauge theories in four dimensions
is a classic and important problem
\ref\thooft{G. 't Hooft, ``A Planar Diagram Theory For Strong Interactions,''
Nucl. Phys. {\bf B72} (1974) 461.}.  The structure of the ``planar diagrams''
that dominate the large $N$ limit gave the first clue that this
problem might be solved by interpreting four-dimensional large $N$ gauge
theory as a string theory.  Attempts in this direction have led to many
insights relevant to critical string theory; for an account of the status,
see  \ref\polyakov{A. M. Polyakov, ``String Theory And Quark Confinement,''
hep-th/9711002.}.
\nref\kleb{S. S. Gubser, I. R. Klebanov, and A. W. Peet, ``Entropy And
Temperature Of Black 3-Branes,'' Phys. Rev. {\bf D54} (1996) 3915.}
\nref\kikleb{I. R. Klebanov, ``World Volume Approach To 
Absorption By Nondilatonic
Branes,'' Nucl. Phys. {\bf B496} (1997) 231.}
\nref\gkt{ S. S. Gubser, I. R. Klebanov,
and A. A. Tseytlin, ``String Theory And Classical Absorption By Threebranes,''
Nucl. Phys. {\bf B499} (1997) 217.}
\nref\jgp{ S. S. Gubser and I. R. Klebanov,
``Absorption By Branes And Schwinger Terms In The World Volume Theory,''
Phys. Lett. {\bf B413} (1997) 41.}
\nref\stro{J. Maldacena and A. Strominger, ``Semiclassical Decay Of
Near Extremal Fivebranes,'' hep-th/9710014.}
Recently, motivated by studies of interactions of
branes with external probes \refs{\kleb  - \stro}, 
and near-extremal brane geometry 
\nref\gibb{G. Gibbons and P. Townsend, ``Vacuum Interpolation In
Supergravity Via Super $p$-Branes,'' Phys. Rev. Lett. {\bf 71} (1993) 5223.}
\nref\dufg{M. J. Duff, G. W. Gibbons, and P. K. Townsend, ``Macroscopic
Superstrings as Interpolating Solitons,'' Phys. Lett. {\bf B332} (1994) 321.}
\nref\fgg{S. Ferrara, G. W. Gibbons, and R. Kallosh, ``Black Holes
And Critical Points In Moduli Space,'' Nucl. Phys. {\bf B500}
(1997) 75, hep-th/9702103; A. Chamseddine, S. Ferrara, G. W. Gibbons,
and R. Kallosh, ``Enhancement Of Supersymmetry Near $5-D$ Black
Hole Horizon,'' Phys. Rev. {\bf D55} (1997) 3647, hep-th/9610155.}
\refs{\gibb,\fgg},  a concrete
proposal in this vein has been made \ref\malda{J. Maldacena,  ``The Large
$N$ Limit Of Superconformal Field Theories And Supergravity,''
hep-th/9711200.}, 
in the context of certain conformally-invariant theories such as $\N=4$
super Yang-Mills theory in four dimensions.  The proposal
relates supergravity on anti-de Sitter or $AdS$ space (or actually
on $AdS$ times a compact manifold) to conformal field theory on the boundary,
and thus potentially introduces into the study of conformal field theory
the whole vast subject of $AdS$ compactification of supergravity
(for a classic review see \ref\duff{M. J. Duff, B. E. W. Nilsson,
and C. N. Pope, ``Kaluza-Klein Supergravity,'' Physics Reports
{\bf 130} (1986) 1.}; see also \ref\oduff{M. J. Duff, H. Lu,
and C. N. Pope, ``$AdS_5\times \S^5$ Untwisted,'' hep-th/9803061.} for
an extensive list of references relevant to current developments).  
Possible relations of a theory on
 $AdS$ space to a theory on the boundary have been explored
for a long time, both in the abstract
(for example, see \ref\flato{M. Flato and C. Fronsdal, ``Quantum Field Theory 
Of
Singletons: The Rac,'' J. Math. Phys. {\bf 22} (1981) 1278; C. Fronsdal,
``The Dirac Supermultiplet,'' Phys. Rev. {\bf D23} (19888) 1982;
E. Angelopoulos, M. Flato, C. Fronsdal, and D. Sternheimer, ``Massless
Particles, Conformal Group, and De Sitter Universe,'' Phys. Rev.
{\bf D23} (1981) 1278.}), and in the context of supergravity and brane
theory (for example, see
\ref\dufo{E. Bergshoeff, M. J. Duff, E. Sezgin, and C. N. Pope,
``Supersymmetric Supermembrane Vacua And Singletons,'' Phys. Lett.
{\bf B199} (1988) 69; M. P. Blencowe and M. J. Duff,
``Supersingletons,'' Phys. Lett. {\bf B203} (1988) 203.}).
More complete references relevant to current developments
can be found in papers already cited and in many of the other important
recent papers
\nref\usfet{S. Hyun, ``$U$-Duality Between Three And Higher Dimensional
Black Holes,'' hep-th/9704005; S. Hyun, Y. Kiem, and H. Shin, ``Infinite
Lorentz Boost Along The $M$ Theory Circle And Nonasymptotically Flat
Solutions In Supergravities,'' hep-th/9712021.}
\nref\sfet{K. Sfetsos and K. Skenderis, ``Microscopic Derivation Of The
Bekenstein-Hawking Formula For Nonextremal Black Holes,'' hep-th/9711138,
H. J. Boonstra, B. Peeters, and K. Skenderis, ``Branes And Anti-de Sitter
Space-times,'' hep-th/9801206.}
\nref\claus{P. Claus, R. Kallosh, and A. van Proeyen, ``$M$ Five-brane
And Superconformal $(0,2)$ Tensor Multiplet In Six-Dimensions,'' 
hep-th/9711161;
P. Claus, R. Kallosh, J. Kumar, P. Townsend, and A. van Proeyen, 
``Conformal Field Theory Of $M2$, $D3$, $M5$, and $D1$-Branes $+$ 
$D5$-Branes,''
hep-th/9801206.}
\nref\kall{R. Kallosh, J. Kumar, and A. Rajaraman, ``Special Conformal
Symmetry Of Worldvolume Actions,'' hep-th/9712073.}
\nref\ferr{S. Ferrara and C. Fronsdal, ``Conformal Maxwell Theory As
A Singleton Field Theory On $AdS(5)$, IIB Three-branes and Duality,''
hep-th/9712239.}
\nref\hyun{S. Hyun, ``The Background Geometry Of DLCQ Supergravity,''
hep-th/9802026.}
\nref\guna{M. Gunaydin and D. Minic, ``Singletons, Doubletons, and $M$
Theory,'' hep-th/9802047.}
\nref\ferf{S. Ferrara and C. Fronsdal, ``Gauge Fields As Composite
Boundary Excitations,'' hep-th/9802126.}
\nref\oog{G. T. Horowitz and H. Ooguri, ``Spectrum Of Large $N$ Gauge Theory
From Supergravity,'' hep-th/9802116.}
\nref\itog{N. Itzhaki, J. M. Maldacena, J. Sonnenschein, and S. Yankielowicz,
``Supergravity And The Large $N$ Limite Of Theories With Sixteen 
Supercharges,''
hep-th/9802042.}
\nref\kach{S. Kachru and E. Silverstein, ``4d Conformal Field Theories
And Strings On Orbifolds,'' hep-th/9802183.}
\nref\berk{M. Berkooz, ``A Supergravity Dual Of A $(1,0)$ Field Theory
In Six Dimensions,'' hep-th/9802195.}
\nref\bala{V. Balasumramanian and F. Larsen, ``Near Horizon Geometry
And Black Holes In Four Dimensions,'' hep-th/9802198.}
\nref\gkp{S. S. Gubser, I. R. Klebanov, and A. M. Polyakov, ``Gauge Theory
Correlators From Noncritical String Theory,'' hep-th/9802109.}
\nref\flato{M. Flato and C. Fronsdal, ``Interacting Singletons,''
hep-th/9803013.}
\nref\nek{A. Lawrence, N. Nekrasov, and C. Vafa, ``On Conformal Theories
In Four Dimensions,'' hep-th/9803015.}
\nref\bersh{M. Bershadsky, Z. Kakushadze, and C. Vafa, ``String Expansion
As Large $N$ Expansion Of Gauge Theories,'' hep-th/9803076.}
\nref\ghk{S. S. Gubser, A. Hashimoto, I. R. Klebanov, and M. Krasnitz,
``Scalar Absorption and the Breaking of the World Volume Conformal 
Invariance,''
hep-th/9803023.}
\nref\aref{I. Ya. Aref'eva and I. V. Volovich, ``On Large $N$ Conformal
THeories, Field Theories On Anti-de Sitter Space, and Singletons,''
hep-th/9803028.}
\nref\casetel{L. Castellani, A. Ceresole, R. D'Auria, S. Ferrara, P. Fr\'e,
and M. Trigiante, ``$G/H$ $M$-Branes And $AdS_{p+2}$ Geometries,''
hep-th/9803039.}
\nref\ferfz{S. Ferrara, C. Fronsdal, and A. Zaffaroni, ``On $N=8$
Supergravity On $AdS_5$ And $N=4$ Superconformal Yang-Mills Theory,''
hep-th/9802203.}
\nref\aoy{O. Aharony, Y. Oz, and Z. yin, hep-th/9803051.}
\nref\minw{S. Minwalla, ``Particles on $AdS_{4/7}$ And Primary Operators
On $M_{2/5}$ Brane Worldvolumes,'' hep-th/9803053.}
\nref\newmalda{J. Maldacena, ``Wilson
Loops In Large $N$ Field Theories,'' hep-th/9803002}
\nref\reyyee{S.-J. Rey and J. Yee, ``Macroscopic Strings As Heavy Quarks
In Large $N$ Gauge Theory And Anti de Sitter Supergravity,'' hep-th/9803001.}
\nref\haly{E. Halyo, ``Supergravity On $AdS_{4/7}\times \S^{7/4}$ and
$M$ Branes,'' hep-th/9803077.}
\nref\leighroz{R. G. Leigh and M. Rozali, ``The Large $N$ Limit Of The
$(2,0)$ Superconformal Field Theory,'' hep-th/9803068.}
\nref\vaffa{M. Bershadsky, Z. Kakushadze, and C. Vafa,
``String Expansion As Large $N$ Expansion Of Gauge Theories,'' hep-th/9803076.}
\nref\rajavin{A. Rajaraman, ``Two-Form Fields And The Gauge Theory
Description Of Black Holes,'' hep-th/9803082.}
\nref\horoross{G. T. Horowitz and S. F. Ross, ``Possible Resolution
Of
Black Hole Singularities From Large $N$ Gauge Theory,'' hep-th/9803085.}
\nref\gomiss{J. Gomis, ``Anti de Sitter Geometry And Strongly
Coupled Gauge Theories,'' hep-th/9803119.}
\nref\fkpz{S. Ferrara, A. Kehagias, H. Partouche, and A. Zaffaroni,
``Membranes And Fivebranes With Lower Supersymmetry And Their
$AdS$ Supergravity Duals,'' hep-th/9803109.}
\nref\mina{J. A. Minahan, ``Quark-Monopole Potentials In Large $N$ Super
Yang-Mills,'' hep-th/9803111.}
\refs{\usfet - \mina}
in which many aspects of the CFT/$AdS$ correspondence have
been extended and better understood.

\nref\witten{E. Witten,  ``Anti-de Sitter Space And Holography,''
hep-th/9802150.}
In \refs{\gkp,\witten}, a precise recipe was presented
for computing CFT observables
in terms of $AdS$ space.
It will be used in the present paper to study in detail a certain
problem in gauge theory dynamics.  The 
problem in question, already discussed in part in section  3.2 of
\witten, is to understand the high temperature behavior
of $\N=4$ super Yang-Mills theory.  As we will see, in this theory, the
CFT/$AdS$ correspondence implies,  in the infinite volume limit,
many expected but subtle quantum
properties, including a non-zero expectation value
for a temporal Wilson loop
\nref\polya{A. M. Polyakov, ``Thermal Properties Of Gauge Fields And
Quark Liberation,'' Phys. Lett. {\bf 72B} (1978) 477.}
\nref\sussk{L. Susskind, ``Lattice Models Of Quark Confinement At High 
Temperature,'' Phys. Rev. {\bf D20} (1979) 2610.}
\refs{\polya,\sussk}, an area law for spatial Wilson loops, 
and a mass gap.  (The study
of Wilson loops is based on a formalism that was introduced recently
\refs{\newmalda,\reyyee}.)
These expectations are perhaps more familiar for ordinary four-dimensional
Yang-Mills
theory without supersymmetry -- for a review see \ref\gross{D. J. Gross,
R. D. Pisarski, and L. G. Yaffe, ``QCD And Instantons At Finite Temperature,''
Rev. Mod. Phys. {\bf 53} (1981) 43.}.  But the incorporation of
supersymmetry, even $\N=4$ supersymmetry, is not expected to affect
these particular issues, since non-zero temperature breaks supersymmetry
explicitly and makes it possible for the spin zero and spin one-half fields
to get mass,\foot{The thermal ensemble on a spatial manifold $\R^3$
can be described by path integrals on $\R^3\times \S^1$, with a radius
for the $\S^1$ equal to  $\beta=T^{-1}$, with $T$ the temperature.  
The fermions obey antiperiodic boundary conditions around the $\S^1$ direction,
and so get masses of order $1/T$ at tree level.  The spin zero bosons
get mass at the one-loop level.} very plausibly reducing the high
temperature behavior to that of the pure gauge theory.  The ability
to recover from the CFT/$AdS$ correspondence
relatively subtle dynamical properties of high temperature gauge theories,
in a situation not governed by supersymmetry or conformal invariance,
certainly illustrates the power of this correspondence.

In section 2, we review the relevant questions about gauge theories
and the framework in which we will work,
and develop a few necessary properties of the Schwarzschild black
hole on $AdS$ space.   The 
CFT/$AdS$ correspondence implies readily that in the limit of large mass,
a Schwarzschild black hole in $AdS$ space has an entropy proportional
to the volume of the horizon, in agreement with the  classic result of
Bekenstein \ref\bekenstein{J. Bekenstein,
``Black Holes And Entropy,'' Phys. Rev. {\bf D7} (1973) 2333.}
and Hawking \ref\hawking{S. W. Hawking, ``Particle Creation By Black
Holes,'' Commun. Math. Phys. {\bf 43} (1975) 199.}.
\nref\horowitz{G. Horowitz and J. Polchinski, ``Corresondence Principle
For Black Holes And Strings,'' Phys. Rev. {\bf D55} (1997) 6189.}
\nref\sussketal{T. Banks, W. Fischler, I. R. Klebanov, and L. Susskind,
``Schwarzschild Black Holes From Matrix Theory,'' Phys. Rev. Lett.
{\bf 80} (1998) 226, hep-th/9709091; I. R. Klebanov and L. Susskind,
``Schwarzschild Black Holes In Various Dimensions From Matrix Theory,''
Phys. Lett. {\bf B416} (1998) 62, hep-th/9719108, ``Schwarzschild
Black Holes In Matrix Theory 2,'' hep-th/9711005.}
\nref\sen{A. Sen, ``Extremal Black Holes And Elementary String States,''
Mod. Phys. Lett. {\bf A10} (1995), hep-th/9504147. }
\nref\vs{A. Strominger and C. Vafa, ``Microscopic Origin Of The 
Bekenstein-Hawking Entropy,'' Phys. Lett. {\bf B379} (1996) 99, 
hep-th/9601029.}
(The comparison of horizon volume of the $AdS$
Schwarzschild solution to field theory entropy was first made, in
the $AdS_5$ case, in \kleb, using a somewhat different language.
As in some other string-theoretic studies of Schwarzschild black holes
\refs{\horowitz,\sussketal}, and some earlier studies of
BPS-saturated black holes \sen, but unlike some microscopic studies
of BPS black holes \vs, in our discussion we
are not able to determine the constant
of proportionality between area and entropy.)  
This way of understanding black hole entropy is in keeping with the notion of
``holography''
\nref\thooftp{G. 't Hooft, ``Dimensional Reduction In Quantum Gravity,''
in {\it Salamfest 93}, p. 284,gr-qc/9310026.}
\nref\thorn{C. Thorn, ``Reformulating String Theory With The $1/N$ Expansion,''
lecture at First A. D. Sakharov Conference on Physics, hep-th/9405069.}
\nref\sussk{L. Susskind, ``The World As A Hologram,'' J. Math. Phys.
{\bf 36} (1995) 6377.} \refs{\thooftp - \sussk}.
The result holds for black holes with Schwarzschild
radius much greater than the radius of curvature
of the $AdS$ space, and so does not immediately imply the corresponding
result for Schwarzschild black holes in Minkowski space.  

In section 3, we demonstrate, on the basis of the CFT/$AdS$ correspondence,
that the $\N=4$ theory at nonzero temperature has the claimed properties,
especially the breaking of the center of the gauge group, magnetic
confinement, and the mass gap.

In section 4, we present, using similar ideas, a proposal
for studying ordinary large $N$ gauge theory in four dimensions
(without supersymmetry or matter fields) via string theory.
In this proposal, we can exhibit confinement and the mass gap, precisely
by the same arguments used in section 3, along with the expected
large $N$ scaling, but we are not able to effectively
compute hadron masses or show that the model is asymptotically free.

\newsec{High Temperatures And $AdS$ Black Holes}

\subsec{$\R^3$ and $\S^3$}

We will study the $\N=4$ theory at finite temperature on a
spatial manifold $\S^3$ or $\R^3$.  $\R^3$ will be obtained by taking
an infinite volume limit starting with $\S^3$.  

To study the theory at finite temperature on $\S^3$, we must
compute the partition function on $\S^3\times \S^1$ -- with 
supersymmetry-breaking boundary conditions in the $\S^1$ directions.
We denote the circumferences of $\S^1$ and $\S^3$ as $\beta$ and $\beta'$,
respectively.  By conformal invariance, only the ratio $\beta/\beta'$ matters.
To study the finite temperature theory on $\R^3$, we  take the large $\beta'$
limit, reducing to $\R^3\times \S^1$.

Once we are on $\R^3\times \S^1$, with circumference $\beta$ for $\S^1$,
the value of $\beta$ can be scaled out via conformal invariance.  Thus,
the $\N=4$ theory on $\R^3$ cannot have a phase transition at any nonzero
temperature.  Even if one breaks conformal invariance by formulating
the theory on $\S^3$ with some circumference $\beta'$, there can be
no phase transition as a function of temperature, since theories
with finitely many local fields have in general no phase transitions
as a function of temperature.  

However, in the large $N$ limit,
it is possible to have phase transitions even in finite volume
\ref\gw{D. J. Gross and E. Witten, ``Possible Third Order Phase
Transition In The Large $N$ Lattice Gauge Theory,'' Phys. Rev.
{\bf D21} (1980) 446.}.  In section 3.2 of \witten,  
it was shown that the $\N=4$ theory on $\S^3\times \S^1$ has in the large
$N$ limit a phase transition as a function of $\beta/\beta'$.  
The large $\beta/\beta'$ phase has some properties in common with
the usual large $\beta$ (or small temperature) phase of confining
gauge theories, while the small $\beta/\beta'$ phase is analogous
to a deconfining phase.  

When we go to $\R^3\times\S^1$ by taking
$\beta'\to\infty$ for fixed $\beta$, we get $\beta/\beta'\to 0$.
So the unique nonzero temperature phase of the $\N=4$ theory on
$\R^3$ is on the high temperature side of the phase transition
and should be compared
to the deconfining phase of gauge theories.  Making this comparison
will be the primary goal of section 3.  We will also make some remarks
in section 3 comparing the low temperature phase on $\S^3\times \S^1$
to the confining phase of ordinary gauge theories.  Here one can make
some suggestive observations, but the scope is limited because in the
particular $\N=4$ gauge theory under investigation, the low temperature
phase on $\S^3$ arises only in finite volume, while most of the deep
questions of statistical mechanics and quantum dynamics
refer to the infinite volume limit.

\subsec{Review Of Gauge Theories}

We will now review the relevant expectations concerning finite
temperature gauge theories in four dimensions.

Deconfinement at high temperatures can be usefully described, in a certain
sense, in terms of spontaneous breaking of the center of the gauge
group (or more precisely of the subgroup of the center under which
all fields transform trivially).  For our purposes, the gauge group
will be $G=SU(N)$, and the center is $\Gamma=\Z_N$; it acts trivially
on all fields, making possible the following standard construction.

Consider  $SU(N)$
gauge theory on $Y\times \S^1$, with $Y$ any spatial manifold.
A conventional gauge transformation is specified by the choice of a map
$g:Y\times \S^1\to G$
which we write explicitly as $g(y,z)$, with $y$ and $z$ denoting
respectively points in $Y$ and in $\S^1$.  (In describing a gauge 
transformation
in this way, we are assuming that the $G$-bundle has been trivialized
at least locally along $Y$; global properties along $Y$ are irrelevant
in the present discussion.)  Such a map has $g(y,z+\beta)=g(y,z)$.
However, as all fields transform trivially under the center of
$G$, we can more generally consider    gauge transformations by
gauge functions $g(y,z)$ that obey 
\eqn\juggle{g(y,z+\beta)=g(y,z)h,}
 with
$h$ an arbitrary element of the center.  Let us call the group of
such extended gauge transformations (with arbitrary dependence on $z$ and $y$
and any $h$) $\bar G$ and the group of ordinary
gauge transformations (with $h=1$ but otherwise unrestricted) 
$G'$.  The quotient $\bar G/G'$ is isomorphic
to the center $\Gamma$ of $G$, and we will denote it simply as $\Gamma$.
Factoring out $G'$ is natural because it acts trivially on all local
observables and physical states (for physical states, $G'$-invariance is 
the statement of Gauss's law), while $\Gamma$ can act nontrivially
on such observables.

An order parameter for spontaneous breaking of $\Gamma$ is the expectation
value of a temporal Wilson line.  Thus, let $C$ be any oriented
closed path of the
form $y\times \S^1$ (with again $y$ a fixed point in $Y$), and consider
the operator
\eqn\ugg{W(C)=\Tr P\exp\int_C A,}
with $A$ the gauge field and
the trace  taken in the $N$-dimensional fundamental representation of $SU(N)$.
Consider a generalized gauge transformation of the form \juggle, with
$h$ an $N^{th}$ root of unity representing an element of the center
of $SU(N)$.  Action by such a gauge transformation multiplies the holonomy
of $A$ around $C$ by $h$, so one has
\eqn\nugg{W(C)\to h W(C).}
Hence, the expectation value $\langle W(C)\rangle$ is an order parameter
for the spontaneous breaking of the $\Gamma$ symmetry.

Of course, such spontaneous symmetry breaking will not occur (for finite
$N$) in finite volume.  But a nonzero expectation value $\langle W(C)\rangle$
in the infinite volume limit, that is with $Y$ replaced by $\R^3$,
is an important order parameter for deconfinement.  Including the Wilson
line $W(C)$ in the system means including an external
 static quark (in the fundamental representation of $SU(N)$), so an expectation
value for $W(C)$ 
means intuitively that the cost in free energy of perturbing the
system by such an external charge
is finite.  In a confining phase, this free energy cost is infinite
and $\langle W(C)\rangle=0$.  The $\N=4$ theory on $\R^3$ corresponds
to a high temperature or deconfining phase; we will confirm in section
3, using the CFT/$AdS$ correspondence, that it has spontaneous breaking of 
the center.  

Other important questions arise if we take the infinite volume
limit, replacing $X$ by $\R^3$.  The theory at long distances
along $\R^3$ is expected to behave like a pure $SU(N)$ gauge theory
in three dimensions.  At nonzero temperature, at least for weak coupling,
the fermions get a mass
at tree level from the thermal boundary conditions in the $\S^1$ direction, 
and the scalars (those present in four dimensions, as well as an extra scalar 
that arises from the component of the gauge field in the $\S^1$ direction)
 get a mass at one loop level; so the long distance dynamics
is very plausibly that of three-dimensional gauge fields.  The main expected 
features of three-dimensional pure Yang-Mills
theory are confinement and a mass gap.  The  mass gap means 
simply that correlation functions $\langle {\cal O}(y,z){\cal O}'(y',z)\rangle$
vanish exponentially for $|y-y'|\to\infty$.  Confinement is expected
to show up in an area law for the expectation value of a spatial Wilson
loop.  The area law means the following.  Let $C$ be now an oriented closed
loop encircling an area $A$ in $\R^3$, at a fixed point on $\S^1$.
The area law means that if $C$ is scaled up, keeping its shape fixed
and increasing $A$, then the expectation value of $W(C)$ vanishes
exponentially with $A$.

\bigskip\noindent{\it ``Confinement'' In Finite Volume}

Finally, there is one more issue that we will address here.
In the large $N$ limit, a criterion for confinement is whether
(after subtracting a constant from the ground state energy)
the free energy is of order one -- reflecting the contributions of
color singlet hadrons -- or of order $N^2$ -- reflecting the contributions
of gluons.  (This criterion has been discussed in \ref\thorn{C. Thorn, 
``Infinite $N(C)$ QCD At Finite Temperature: Is There An Ultimate 
Temperature?'' Phys. Lett. {\bf B} (1981) 458.}.)
In \witten, it was shown that in the $\N=4$ theory on $\S^3\times \S^1$,
the large $N$ theory has a low temperature phase
with a free energy of order 1 -- a ``confining'' phase -- and a high
temperature phase with a free energy of order $N^2$ -- an
``unconfining'' phase.

Unconfinement at high temperatures comes as no surprise, 
of course, in this theory, and since the
theory on $\R^3\times \S^1$, at any temperature, is in the high temperature
phase, we recover the expected result that the infinite volume theory is not
confining.  However, it seems strange that the finite volume theory on $\S^3$,
at low temperatures, is ``confining'' according to this particular
criterion.

This, however, is a general property of large $N$ gauge theories on $\S^3$,
at least for weak coupling (and presumably also for strong coupling).
On a round three-sphere, the classical solution of lowest energy
is unique up to gauge transformations (flat directions in the scalar potential
are eliminated by the $R\phi^2$ coupling to scalars, $R$ being the Ricci
scalar), and is given by setting the
gauge field $A$, fermions $\psi$, and scalars $\phi$ all to zero.
This configuration is invariant under global $SU(N)$ gauge transformations.
The Gauss law constraint in finite volume says that physical states
must be invariant under the global $SU(N)$.
There are no zero modes for any fields (for scalars this statement
depends on the $R\phi^2$ coupling).
Low-lying excitations are obtained by acting on the ground state with a finite
number of $A$, $\psi$, and $\phi$ creation operators, and then imposing
the constraint of global $SU(N)$ gauge invariance.  The creation operators
all transform in the adjoint representation, and so are represented in
color space by matrices $M_1,M_2,\dots,M_s$.   $SU(N)$ invariants are
constructed as traces, say $\Tr\,M_1M_2\dots M_s$.  The number of such
traces is given by the number of ways to order the factors and is independent
of $N$.  So the multiplicity of low energy states is independent of $N$,
as is therefore the low temperature free energy.

This result, in particular, is kinematic, and has nothing to do with
confinement.  To see confinement from the $N$ dependence of the free energy,
we must go to infinite volume.  On $\R^3$, 
the Gauss law constraint does not say that the physical states are
invariant under global $SU(N)$ transformations, but only that their global
charge is related to the electric field at spatial infinity.  If the free
energy on $\R^3$ is of order 1 (and not proportional to $N^2$), this
actually is an order parameter for confinement.

Now let us go back to finite volume and consider the behavior at
high temperatures.  At high temperatures, one 
cannot effectively compute the free energy
by counting elementary excitations.  It is more efficient
to work in the ``crossed channel.'' In $\S^3\times \S^1$, with circumferences
$\beta'$ and $\beta$, if we take $\beta'\to\infty$ with fixed $\beta$,
the free energy is proportional to the volume of $\S^3$ times the ground
state energy density of the $2+1$-dimensional theory that is obtained by
compactification on $\S^1$ (with circumference $\beta$ and 
supersymmetry-breaking boundary conditions).
That free energy is of order $N^2$ (the supersymmetry breaking spoils the
cancellation between bosons and fermions already at the one-loop level, and the
one-loop contribution is proportional to $N^2$).
The volume of $\S^3$ is of order $(\beta')^3$. So the free energy
on $\S^3\times \S^1$ scales as $N^2(\beta')^3$ if one takes $\beta'\to\infty$
at fixed $\beta$, or in other words as $N^2\beta^{-3}$ if one takes
$\beta\to 0$ at fixed $\beta'$.   Presently we will recover this
dependence on $\beta$ by comparing to black holes.

\subsec{$AdS$ Correspondence And Schwarzshild Black Holes}

The version of the CFT/$AdS$ correspondence that we will use
asserts that conformal field theory on an $n$-manifold $M$
is to be studied by summing over contributions of Einstein manifolds
$B$ of dimension $n+1$
which (in a sense explained in \refs{\gkp,\witten}) have
$M$ at infinity.

\def\B{{\bf B}}
We will be mainly interested in the case that $M=\S^{n-1}\times \S^1$,
or $\R^{n-1}\times \S^1$.  For $\S^{n-1}\times \S^1$, there are
two known $B$'s, identified by Hawking and Page 
\ref\hp{S. W. Hawking and D. Page,
``Thermodynamics Of Black Holes In Anti-de Sitter Space,''
Commun. Math. Phys. {\bf 87} (1983) 577.} 
in the context of quantum gravity on $AdS$ space.
One manifold, $X_1$, is the quotient of $AdS$ space by a subgroup
of $SO(1,n+1)$ that is isomorphic to $\Z$.  The metric (with
Euclidean signature) can be written 
\eqn\nuggo{ds^2=\left({r^2\over b^2}+1\right)dt^2+{dr^2\over 
\left({r^2\over b^2}\right)
+1}+r^2d\Omega^2,}
with $d\Omega^2$ the metric of a round sphere $\S^{n-1}$ of unit
radius.  
Here $t$ is a periodic variable of arbitrary period.
We have normalized \nuggo\ so that the Einstein equations read
\eqn\theydoread{R_{ij}=-nb^{-2}g_{ij};}
here $b$ is the radius of curvature of the anti-de Sitter
space.
With this choice, $n$ does not appear explicitly in the  metric.
This manifold can contribute to either the standard thermal
ensemble $\Tr e^{-\beta H}$ or to $\Tr (-1)^Fe^{-\beta H}$, depending
on the boundary conditions one uses for fermions in the $t$ direction.
The topology of $X_1$ is $\R^n\times \S^1$, or $\B^n\times \S^1$
($\B^n$ denoting an $n$-ball) if we compactify it by including the boundary
points at $r=\infty$.

The second solution, $X_2$, is the Schwarzschild black hole, in $AdS$
space.  The metric is
\eqn\puggo{ds^2=\left({r^2\over b^2}+1-{w_nM\over r^{n-2}}\right)dt^2+
{dr^2\over \left({r^2\over b^2}+1-{w_nM\over r^{n-2}}\right)}+r^2d\Omega^2.}
Here $w_n$ is the constant
\eqn\furry{w_n={16\pi G_N\over (n-1){\rm Vol}(\S^{n-1})}.}
Here $G_N$ is the $n+1$-dimensional Newton's constant
and ${\rm Vol}(\S^{n-1})$ is the volume of a unit $n-1$-sphere; 
the factor $w_n$
is included so that $M$ is the mass of the black hole (as we will compute
later).  Also, the spacetime is
 restricted to the region $r\geq r_+$, with $r_+$ the largest
solution of the equation
\eqn\vuggo{{r^2\over b^2}+1-{w_nM\over r^{n-2}}=0.}

The metric \puggo\ is smooth and complete if and only if the period of
$t$ is
\eqn\hulfo{\beta_0={4\pi b^2r_+\over nr_+^2+(n-2)b^2}.}
For future use, note that in the limit of large $M$ one has
\eqn\ulgo{\beta_0\sim {4\pi b^2\over n(w_nb^2)^{1/n}M^{1/n}}.}
As in the $n=3$ case considered in \hp, $\beta_0$ has a maximum as
a function of $r_+$, so the Schwarzschild black hole only contributes
to the thermodynamics if $\beta$ is small enough, that is if the temperature
is high enough.  Moreover, $X_2$ makes the dominant contribution
at sufficiently high temperature, while $X_1$ dominates at low temperature.
The topology of $X_2$ is $\R^2\times 
\S^{n-1}$, or $\B^2\times \S^{n-1}$ if we compactify it to include boundary
points.  In particular, $X_2$ is simply-connected,  has a unique
spin structure, and contributes to the standard thermal ensemble but
not to $\Tr (-1)^F e^{-\beta H}$. 

With either \nuggo\ or \puggo, the geometry of the $\S^{n-1}\times \S^1$  
factor at large $r$ can be simply explained: the $\S^1$ has radius
approximately $\beta=(r/b)\beta_0$, and the $\S^{n-1}$ has radius 
$\beta'=r/b$.  The ratio is thus $\beta/\beta'=\beta_0$.  If we wish
to go to $\S^1\times \R^{n-1}$, we must take $\beta/\beta'\to 0$, that
is $\beta_0\to 0$; this is the limit of large temperatures.
   \hulfo\ seems to show that this can be done with
either $r_+\to 0$ or $r_+\to \infty$, but the $r_+\to 0$ branch
is thermodynamically unfavored \hp\ (having  larger action), so we must
take the large $r_+$ branch, corresponding to large $M$.

A scaling that reduces \hulfo\ to a solution with boundary
$\R^{n-1}\times \S^1$ may be made as follows.  If we set 
$r=(w_nM/b^{n-2})^{1/n}\rho$,
$t=(w_nM/b^{n-2})^{-1/n}\tau$, 
then for large $M$ we can reduce $r^2/b^2+1-w_nM/r^{n-2}$
to $(w_nM/b^{n-2})^{2/n}(\rho^2/b^2-b^{n-2}/\rho^{n-2})$.  
The period of $\tau$ become $\beta_1=(w_nM/b^{n-2})^{1/n}\beta_0$ or 
(from \ulgo) for large $M$
\eqn\coofball{\beta_1={4\pi b\over n}.}
The metric becomes
\eqn\goofball{ds^2=\left({\rho^2\over b^2}-
{b^{n-2}\over\rho^{n-2}}\right)d\tau^2
+{d\rho^2\over {\rho^2\over b^2}-{b^{n-2}\over\rho^{n-2}}} +
(w_nM/b^{n-2})^{2/n}\rho^2\,
d\Omega^2.}
The $M^{2/n}$ multiplying the last term means that the radius of 
$\S^{n-1}$  is of order $M^{1/n}$ and so diverges for $M\to\infty$.
Hence, the $\S^{n-1}$ is becoming flat and looks for $M\to\infty$ locally
like $\R^{n-1}$.  If we introduce near a point $P\in\S^{n-1}$ coordinates
$y_i$ such that at $P$, $d\Omega^2=\sum_i dy_i^2$, and then set
$y_i=(w_nM/b^{n-2})^{-1/n}x_i$, then the metric becomes
\eqn\oofball{ds^2=\left({\rho^2\over b^2}-{b^{n-2}\over\rho^{n-2}}\right)d\tau^
2
+{d\rho^2\over \left({\rho^2\over b^2}-{b^{n-2}\over\rho^{n-2}}\right)} +
\rho^2\sum_{i=1}^{n-1}dx_i^2.}
This is the desired solution $\tilde X$ that is asymptotic at infinity to
$\R^{n-1}\times \S^1$ instead of $\S^{n-1}\times\S^1$.
Its topology, if we include boundary points, is $\R^{n-1}\times \B^2$.
The same solution was found recently by scaling of a near-extremal
brane solution \horoross.

\subsec{Entropy Of Schwarzschild Black Holes}

Following Hawking and Page \hp\ (who considered the case $n=3$), we will
now describe the thermodynamics of Schwarzschild black holes in $AdS_{n+1}$.
Our normalization of the cosmological constant is stated in \theydoread.
The bulk Einstein action with this
value of the cosmological constant is
\eqn\hhp{I=-{1\over 16 \pi G_N}\int d^{n+1}x\sqrt g\left(R+{\half n(n-1)\over
b^2}\right).}
For a solution of the equations of motion, one has $R=-\half n(n+1)/b^2$, and
the action becomes
\eqn\nhp{I={n\over 8\pi G_N}\int d^{n+1}x\sqrt g,}
that is, the volume of spacetime times $n/8\pi G_N$.  The action additionally
has a surface term 
\nref\york{J. W. York, ``Role Of Conformal Three-Geometry In The
Dynamics Of Gravitation,'' Phys. Rev. Lett. {\bf 28} (1972) 1082.}
\nref\gib{G. W. Gibbons and S. W. Hawking, ``Action Integrals And
Partition Functions In Quantum Gravity,'' Phys. Rev. {\bf D15} (1977) 2752.}
\refs{\york,\gib}, but the surface term vanishes for the $AdS$ Schwarzschild
black hole, as noted in \hp, because the black hole correction to the $AdS$
metric vanishes too rapidly at infinity.

Actually, both the $AdS$ spacetime \nuggo\ and the black hole spacetime
\puggo\ have infinite volume.  As in \hp, one subtracts the two volumes
to get a finite result.  Putting an upper cutoff $R$ on the radial
integrations, the regularized volume of the $AdS$ spacetime
is
\eqn\ombo{V_1(R)=\int_0^{\beta'} dt\int_0^R dr\int_{\S^{n-1}} 
d\Omega \,\,r^{n-1},}
and the regularized volume of the black hole spacetime is
\eqn\ombo{V_2(R)=\int_0^{\beta_0} dt\int_{r_+}^R dr\int_{\S^{n-1}}d\Omega
 \,\,r^{n-1}.}
One difference between the two integrals is obvious here: in the black
hole spacetime $r\geq r_+$, while in the $AdS$ spacetime $r\geq 0$.
A second and slightly more subtle difference is that one must use
different periodicities $\beta'$ and $\beta_0$ for the $t$ integrals in the
two cases.  The black hole spacetime is smooth only if $\beta_0$ has
the value given in \hulfo, but for the $AdS$ spacetime, any value of
$\beta'$ is possible.  One must adjust $\beta'$ so that the geometry
of the hypersurface $r=R$ is the same in the two cases; this is done by
setting $\beta'\sqrt{(r^2/b^2)+1}=\beta_0\sqrt{(r^2/b^2)+1-w_nM/r^{n-2}}.$
After doing so, one finds that the action difference is
\eqn\muvv{I={n\over 8\pi G_N}\lim_{R\to\infty}(V_2(R)-V_1(R))
={{\rm Vol}(\S^{n-1})(b^2r_+^{n-1}-r_+^{n+1})\over 4G_N(nr_+^2+(n-2)b^2)}.}
This is positive for small $r_+$ and negative for large $r_+$, showing
that the phase transition found in \hp\ occurs for all $n$.

Then, as in \hp, one computes the energy
\eqn\imopo{E={\partial I\over \partial\beta_0}=
{(n-1){\rm Vol}(\S^{n-1})(r_+^nb^{-2}+r_+^{n-2})\over 16\pi G_N}= M}
and the entropy
\eqn\ploopo{S=\beta_0E-I={1\over 4G_N}r_+^{n-1}{\rm Vol}(\S^{n-1})}
of the black hole.  The entropy can be written
\eqn\loopo{S={A\over 4G_N},}
with $A$ the volume of the horizon, which is the surface at $r=r_+$.

\bigskip\noindent{\it Comparison To Conformal Field Theory}

Now we can compare this result for the entropy to the predictions of
conformal field theory.

The black hole entropy should be compared to boundary conformal field
theory on $\S^{n-1}\times \S^1$, where the two factors have
circumference 1 and $\beta_0/b$, respectively.  In the limit as $\beta_0\to 0$,
this can be regarded as a high temperature system on $\S^{n-1}$. Conformal
invariance implies that the entropy density on $\S^{n-1}$
scales, in the limit of small $\beta_0$,
as $\beta_0^{-(n-1)}$.  According to \hulfo, $\beta_0\to 0$ means
$r_+\to\infty$ with $\beta_0\sim 1/r_+$.  Hence, the boundary conformal
field theory predicts that the entropy of this system is of order
$r_+^{n-1}$, and thus asymptotically is a fixed multiple of the horizon
volume which appears in \loopo.  This is of course the classic
result of Bekenstein and Hawking, for which microscopic explanations
have begun to appear only recently.
 Note that this discussion assumes
that $\beta_0<<1$, which means that $r_+>>b$; so it applies only
to black holes whose Schwarzschild radius is much greater than the radius
of curvature of $AdS$ space.  However, in this limit, one does get
a simple explanation of why the black hole entropy is proportional to area.
The explanation is entirely ``holographic'' in spirit \refs{\thooftp,\sussk}.

\nref\cardy{J. Cardy, ``Operator Content Of Two-Dimensional
Conformally-Invariant Theories,'' Nucl. Phys. {\bf B270} (1986) 186.}
\nref\sstrominger{A. Strominger, ``Black Hole Entropy From Near-Norizon
Microstates,'' hep-th/9712251.}
To fix the constant of proportionality between entropy and horizon volume
(even in the limit of large black holes), one needs some additional general
insight, or some knowledge of the quantum field theory on the boundary.
For $2+1$-dimensional black holes, in the context of an old framework
\ref\brh{J. D. Brown and M. Henneaux, ``Central Charges In The Canonical
Realization Of Asymptotic Symmetries: An Example From Three-DImensional
Gravity,'' Commun. Math. Phys. {\bf 104} (1986) 207.}
for a relation to boundary conformal field theory which actually
is a special case of the general CFT/$AdS$ correspondence,
such additional information is provided
by modular invariance of the boundary conformal field theory
\refs{\cardy,\sstrominger}.

\newsec{High Temperature Behavior Of The $\N=4$ Theory}

In this section we will address three questions about the high
temperature behavior of the $\N=4$ theory that were raised in section 2:
the behavior of temporal Wilson lines; the behavior of spatial Wilson
lines; and the existence of a mass gap.

In discussing Wilson lines, we use a formalism proposed recently
\refs{\newmalda,\reyyee}.  
Suppose one is doing physics on a four-manifold $M$ which
is the boundary of a five-dimensional Einstein manifold $B$ (of negative
curvature).  To compute a Wilson line associated with a contour $C\subset M$,
we study elementary strings on  $B$ with the property that
the string worldsheet $D$ has $C$ for its boundary.  Such a $D$ has
an infinite area, but the divergence is proportional to the circumference of
$C$.  One can define therefore a regularized area $\alpha(D)$ by
subtracting from the area of $D$ an infinite multiple of the circumference of
$C$.  The expectation value of a Wilson loop $W(C)$ is then roughly
\eqn\totally{\langle W(C)\rangle =\int_{\cal D} d\mu \,e^{-\alpha(D)}}
where ${\cal D}$ is the space of string worldsheets obeying the boundary
conditions and $d\mu$ is the measure of the worldsheet path integral.
Moreover, according to \refs{\newmalda,\reyyee}, 
in the regime in which supergravity is valid
(large $N$ and large $g^2N$), the integral can be evaluated approximately
by setting $D$ to the surface of smallest $\alpha(D)$ that obeys the
boundary conditions.

The formula \totally\
is oversimplified for various reasons.  For one thing,
worldsheet fermions must be included in the path integral.  
Also, the description
of the $\N=4$ theory actually involves not strings on $B$ but strings on the
ten-manifold $B\times \S^5$.  Accordingly, what are considered in
\refs{\newmalda,\reyyee} are
some generalized Wilson loop operators with scalar fields included in the
definition; the boundary behavior of $D$ in the
$\S^5$ factor depends on which operator one uses.  But if all scalars
have masses, as they do in the $\N=4$ theory at positive temperature,
the generalized Wilson loop operators are equivalent at long distances
to conventional ones.
An important conclusion from \totally\ nonetheless stands: Wilson
loops on $\R^4$ will obey an area law if, when $C$ is scaled up, the minimum
value of $\alpha(D)$ scales like a positive multiple of the area enclosed
by $C$.   \totally\ also implies vanishing of $\langle W(C)\rangle$ if
suitable $D$'s do not exist, that is, if $C$ is not a boundary in $B$.

\subsec{Temporal Wilson Lines}

Our first goal will be to analyze temporal Wilson lines.  That is, we take
spacetime to be $\S^3\times \S^1$ or $\R^3\times \S^1$, and we take
$C=P\times \S^1$, with $P$ a point in $\S^3$ or in $\R^3$.

We begin on $\S^3$ in the low temperature phase.  We recall that
this is governed by a manifold $X_1$ with the topology of $\B^4\times \S^1$.
In particular, the contour $C$, which wraps around the $\S^1$, is not
homotopic to zero in $X_1$ and is not the boundary of any $D$.
Thus, the expectation value of a temporal Wilson line vanishes at low
temperatures.  This is the expected result, corresponding to the fact
that the center $\Gamma$ of the gauge group is unbroken at low
temperatures.

Now we move on to the high temperature phase on $\S^3$.  This phase
is governed by a manifold $X_2$ that is topologically $\S^3\times \B^2$.
In this phase, $C=P\times \S^1$ is a boundary; in fact it is the boundary
of $D=P\times \B^2$.  Thus, it appears at first sight that the temporal
Wilson line has a vacuum expectation value and that the center of the
gauge group is spontaneously broken.

There is a problem here.  Though we expect these results in the
high temperature phase on $\R^3$, they cannot hold on $\S^3$, because
the center (or any other bosonic symmetry) cannot be spontaneously broken
in finite volume.  The resolution of the puzzle is instructive.
The classical solution on $X_2$ is not unique.  We must recall that Type IIB
superstring theory has a two-form field $B$ that couples to the elementary
string world-sheet $D$ by
\eqn\ikno{i\int_D B.}
The gauge-invariant field strength is $H=dB$.  We can add to the solution
a ``world-sheet theta angle,'' that is a $B$ field of $H=0$ with an
arbitrary value of $\psi=\int_{D}B$ (here $D$ is any surface obeying
the boundary conditions, for instance $D=P\times \B^2$).  Since discrete gauge
transformations that shift the flux of $B$ by a multiple of $2\pi$ are
present in the theory, $\psi$ is an angular variable with period $2\pi$.

If this term is included, the path integrand in \totally\ receives
an extra factor $e^{i\psi}$.  Upon integrating over the space of all
classical solutions -- that is integrating over the value of $\psi$ -- the
expectation value of the temporal Wilson line on $\S^3$ vanishes.

Now, let us go to $\R^3\times \S^1$, which is the boundary of
$\R^3\times \B^2$.  In infinite volume, $\psi$ is best understood
as a massless scalar field in the low energy theory on $\R^3$.  
One still integrates over 
local fluctuations in $\psi$, but not over the vacuum expectation value of
$\psi$, which is set by the value at spatial infinity.  The expectation
value of $W(C)$ is nonzero and is proportional to $e^{i\psi}$.

What we have seen is thus spontaneous symmetry breaking: in infinite
volume, the expectation value of $\langle W(C)\rangle$ is nonzero,
and depends on the choice of vacuum, that is on the value of $\psi$.
The field theory analysis that we reviewed in section 2 indicates that
the symmetry that is spontaneously broken by the choice of $\psi$ is the
center, $\Gamma$, of the gauge group.  Since $\psi$ is a continuous angular
variable, it seems that the center is $U(1)$.  This seems to imply that
the gauge group is not $SU(N)$, with center $\Z_N$, but $U(N)$.  However,
a variety of arguments \witten\ show that the $AdS$ theory encodes
a $SU(N)$ gauge group, not $U(N)$.  Perhaps the apparent $U(1)$ center
should be understood as a large $N$ limit of $\Z_N$.

\bigskip\noindent{\it 't Hooft Loops}

We would also like to consider in a similar way 't Hooft loops.
These are obtained from Wilson loops by electric-magnetic duality.
Electric-magnetic duality of $\N=4$ arises 
\nref\green{M. B. Green and M. Gutperle, ``Comments On Three-Branes,''
Phys. Lett. {\bf B377} (1996) 28.}\nref\tseytlin{A. A. Tseytlin, 
``Self-duality of Born-Infeld Action
And Dirichlet 3-Brane Of Type IIB Superstring,'' Nucl. Phys. {\bf B469}
(1996) 51.}\refs{\green,\tseytlin} directly from the $\tau\to -1/\tau$
symmetry of Type IIB.  That symmetry
exchanges elementary strings with $D$-strings.
So to study the 't Hooft loops we need only, as in \mina,
replace elementary strings
by $D$-strings in the above discussion.    

The $\tau\to-1/\tau$ symmetry
exchanges the Neveu-Schwarz two-form $B$ which entered the
above discussion with its Ramond-Ramond counterpart $B'$;
the $D$-brane theta angle $\psi'=\int_{D}B'$ thus plays the role
of $\psi$ in the previous discussion.  In the thermal physics on
$\R^3\times\S^1$, the center of the ``magnetic gauge group'' is spontaneously
broken, and the temporal 't Hooft loops have an expectation value,
just as we described for Wilson loops.  The remarks that we make
presently about spatial Wilson loops similarly carry over for spatial 't
Hooft loops.

\subsec{Spatial Wilson Loops}

Now we will investigate the question of whether at nonzero temperature
the spatial Wilson loops obey an area law.
The main point is to first understand why
there is {\it not} an area law at zero temperature.
At zero temperature, one works with the $AdS$ metric
\eqn\ivvo{ds^2={1\over x_0^2}\left(dx_0^2+\sum_{i=1}^4dx_i^2\right).}
We identify the spacetime $M$ of the $\N=4$ theory with the boundary at 
$x_0=0$,
 parametrized by the
Euclidean coordinates $x_i$, $i=1,\dots, 4$.  $M$ has a metric
$d\tilde s^2=\sum_idx_i^2$ obtained by multiplying $ds^2$ by $x_0^2$
and setting $x_0=0$. (If we use a function other than $x_0^2$, the metric
on $M$ changes by a conformal transformation.) We take a closed oriented
curve $C\subset M$ and regard it as the boundary of an oriented compact
surface $D$ in $AdS$ space.  The area of $D$ is infinite, but after subtracting
an infinite counterterm proportional to the circumference of $C$,
we get a regularized area $\alpha(D)$.  In the framework of \refs{\newmalda,
\reyyee}, the expectation value of the Wilson
line $W(C)$ is proportional to $\exp(-\alpha(D))$, with $D$ chosen to
minimize $\alpha(D)$.
 
Now the question arises: why does not this formalism {\it always} give
an area law?  As the area enclosed by $C$ on the boundary
is scaled up, why is not the area of
$D$ scaled up proportionately?  The answer to this is clear from conformal
invariance.  If we scale up $C$ via $x_i\to tx_i$, with large positive $t$,
then by conformal invariance we can scale up $D$, with $x_i\to t x_i$, 
$x_0\to t x_0$, without changing its area (except for a boundary term
involving the regularization).
Thus the area of $D$ need not be proportional to the area enclosed by
$C$ on the boundary.  Since, however, in this process
we had to scale $x_0\to tx_0$ with very large $t$,
the surface $D$ which is bounded by a very large circle $C$ ``bends'' very
far away from the boundary of $AdS$ space.  If such a bending of $D$ were 
prevented -- if $D$ were limited to a region with $x_0\leq L$ for some cutoff
$L$ -- then one would get an area law for $W(C)$.  This is precisely what
will happen at nonzero temperature.

At nonzero temperature, we have in fact the metric \oofball\ obtained 
earlier, with $n=4$:
\eqn\ploofball{ds^2=\left({\rho^2\over b^2}-{b^2\over\rho^{2}}\right)d\tau^2
+{d\rho^2\over \left({\rho^2\over b^2}-{b^2\over\rho^{2}}\right)} +
\rho^2\sum_{i=1}^{3}dx_i^2.}
The range of $\rho$ is $b\leq \rho\leq \infty$.
Spacetime -- a copy of $\R^3\times \S^1$ -- is the boundary at $\rho=\infty$.
We define a metric on spacetime by dividing by $\rho^2$ and setting
$\rho=\infty$.  In this way we obtain the spacetime metric
\eqn\bpob{d\tilde s^2={d\tau^2\over b^2}+\sum_{i=1}^3dx_i^2.}
As the period of $\tau$ is $\beta_1$, the circumference of the $\S^1$ factor
in $\R^3\times \S^1$ 
is $\beta_1/b$ and the temperature is 
\eqn\homey{T={b\over \beta_1}={1\over \pi}.}
Because of conformal invariance, the numerical value of course does not matter.

Now, let $C$ be a Wilson loop in $\R^3$, at a fixed value of $\tau$, enclosing
an area $A$ in $\R^3$.  A bounding surface $D$ in the spacetime \ploofball\
is limited to $\rho\geq b$, so the coefficient of $\sum_idx_i^2$
is always at least $b^2$.  Apart from a surface term that depends
on the regularization and the detailed solution of the equation for a minimal
surface, the regularized area of $D$ is at least $\alpha(D)=b^2A$
(and need be no larger than this).  The Wilson loops therefore obey an
area law, with string tension $b^2$ times the elementary
Type IIB string tension.

We could of course have used a function other than $\rho^2$
in defining the spacetime metric, giving a conformally equivalent
metric on spacetime.  For instance,  picking a constant $s$ and using
$s^2\rho^2$ instead of $\rho^2$ would scale the temperature as $T\to T/s$
and would multiply all lengths on $\R^3$ by $s$.  The area enclosed by $C$
would thus become $A'=As^2$.  As $\alpha(D)$ is unaffected, the relation
betseen $\alpha(D)$ and $A'$ becomes $\alpha(D)=\left(b^2/s^2\right)
\cdot A'$.  The string tension in the Wilson loop area law thus scales like
$s^{-2}$, that is, like $T^2$, as expected from conformal invariance.

\subsec{The Mass Gap}

The last issue concerning the $\N=4$ theory at high temperature that
we will discuss here is the question of whether there is a mass gap.
We could do this by analyzing correlation functions, using the formulation
of \refs{\gkp,\witten}, but it is more direct to use a Hamiltonian approach
(discussed  at the end of \witten) in which one identifies
the quantum states of the supergravity theory with those of the quantum
field theory on the boundary.  

\def\SSigma{{\bf W}}
\nref\isham{S. J. Avis, C. Isham, and D. Storey, ``Quantum Field Theory
In Anti-de Sitter Space-time,'' Phys. Rev. {\bf D18} (1978) 3565.}
\nref\bfud{P. Breitenlohner and D. Z. Freedman, ``Positive Energy In
Anti-de Sitter Backgrounds And Gauged Extended Supergravity,''
Phys. Lett. {\bf 115B} (1982) 197, ``Stability In Gauged Extended
Supergravity,'' Ann. Phys. {\bf 144} (1982) 197.}
So we will demonstrate a mass gap by showing that there is a gap,
in the three-dimensional sense, for quantum fields propagating on
the five-dimensional spacetime
\eqn\poofball{ds^2=\left({\rho^2\over b^2}-{b^2\over\rho^{2}}\right)d\tau^2
+{d\rho^2\over \left({\rho^2\over b^2}-{b^2\over\rho^{2}}\right)} +
\rho^2\sum_{i=1}^{3}dx_i^2.}
 This spacetime is the product of a three-space
$\R^3$, parametrized by the $x_i$, with a two-dimensional ``internal space''
$\SSigma$, parametrized by $\rho$ and $\tau$.
We want to show that a quantum free field propagating on this five-dimensional
spacetime gives rise, in the three-dimensional sense, to a discrete spectrum
of particle masses, all of which are positive.  When such a spectrum is
perturbed by interactions, the discreteness of the spectrum is lost (as
the very massive particles become unstable), but the mass gap persists.

If $\SSigma$ were compact, then  discreteness of the mass spectrum
would be clear: particle masses on $\R^3$ would arise from eigenvalues
of the Laplacian (and other wave operators) on $\SSigma$.  Since $\SSigma$
is not compact, it is at first sight surprising that a discrete mass
spectrum will emerge.  However, this does occur, by essentially the same
mechanism that leads to discreteness of particle energy levels
on $AdS$ space \refs{\isham,\bfud} with a certain notion of energy.

For illustrative purposes, we will consider the propagation of a Type IIB
dilaton field $\phi$ on this spacetime.  Other cases are similar.
The action for $\phi$ is
\eqn\behbu{\eqalign{I(\phi)= &\half
\int_b^\infty d\rho\int_0^{\beta_1/b}d\tau
\int_{-\infty}^\infty d^3x\,\,\,\rho^3\left(\left({\rho^2\over b^2}-{b^2\over
\rho^{2}}\right)
\left({\partial\phi\over \partial\rho}\right)^2 \right.\cr &\left.
+\left({\rho^2\over b^2}-{b^2\over
\rho^{2}}\right)^{-1}\left({\partial\phi\over \partial\tau}\right)^2
+\rho^{-2}\sum_i\left({\partial\phi\over\partial x_i}\right)^2\right).\cr}}
Since translation of $\tau$ is a symmetry, modes with different momentum
in the $\tau$ direction are decoupled from one another. The spectrum of
such momenta is discrete (as $\tau$ is a periodic variable). To simplify
things slightly and illustrate the essential point, we will write
the formulas for the modes that are independent of $\tau$; others simply
give, by the same argument,
 additional three-dimensional massive particles with larger masses.

We look for a solution of the form $\phi(\rho,x)=f(\rho)e^{ik\cdot x}$,
with $\vec k$ the momentum in $\R^3$.  The effective Lagrangian becomes
\eqn\ehbu{I(f)=\half
\int_b^\infty d\rho\,\,\rho^3\left((\rho^2/b^2-b^2/\rho^{2})
\left({df\over d\rho}\right)^2
+\rho^{-2}k^2f^2\right).}
The equation of motion for $f$ is 
\eqn\bhu{-\rho^{-1}{d\over d\rho}\left(\rho^3(\rho^2/b^2-b^2/\rho^{2}){df\over
d\rho}\right)+ k^2 f=0.}
A mode of momentum $k$ has a mass $m$, in the three-dimensional sense, that
is given by $m^2=-k^2$.
We want to show that the equation \bhu\ has acceptable solutions only
if $m^2$ is in a certain discrete set of positive numbers.

Acceptable solutions are those that obey the following boundary conditions:

(1) At the lower endpoint $\rho=b$, we require $df/d\rho=0$.
The reason for this is that $\rho$ behaves near this endpoint as the origin
in polar coordinates; hence $f$ is not smooth at this endpoint unless
$df/d\rho=0$ there.  

(2) For $\rho\to\infty$, the equation has two linearly independent
solutions, which behave as $f\sim {\rm constant}$ and $f\sim \rho^{-4}$.
  We want a normalizable solution, so we require
that $f\sim \rho^{-4}$.    

For given $k^2$, the equation \bhu\ has, up to a constant multiple, a unique
solution that obeys the correct boundary condition near the lower endpoint.
For generic $k^2$, this solution will approach a nonzero
constant for $\rho\to\infty$.
As in standard quantum mechanical problems, there is a normalizable
solution only if $k^2$ is such that the solution
that behaves correctly at the lower endpoint also vanishes for $\rho\to\infty$.
This ``eigenvalue'' condition determines a discrete set of values of $k^2$.

The spectrum thus consists entirely of a discrete set of normalizable
solutions.  There are no such normalizable solutions for $k^2\geq 0$.
This can be proved by noting that, given a normalizable solution $f$ of
the equation of motion, a simple integration by parts shows that the action
\ehbu\ vanishes.  For $k^2\geq 0$, vanishing of $I(f)$ implies that
$df/d\rho=0$, whence (given normalizability) $f=0$.  So the discrete
set of values of $k^2$ at which there are normalizable solutions are all 
negative; the masses $m^2=-k^2$ are hence strictly positive.  This
confirms the existence of the mass gap.

To understand the phenomenon better, let us compare to what usually
happens in quantum mechanics.
In typical quantum mechanical scattering problems, with potentials
that vanish at infinity, the solutions with
positive energy (analogous to $m^2>0$)
are oscillatory at infinity and obey plane wave normalizability.
When this is so, both solutions at infinity  are physically acceptable
(in some situations, for example, they are interpreted as incoming
and outgoing waves),
and one gets a continuous spectrum that starts at zero energy.  
The special property of the
problem we have just examined is that even for negative $k^2$, there
are no oscillatory solutions at infinity, and instead one of the two
solutions must be rejected as being unnormalizable near $\rho=\infty$.
This feature leads to the discrete spectrum.

If instead of the spacetime
\poofball, we work on $AdS$ spacetime \nuggo, there is a continuous
spectrum of solutions
with plane wave normalizability for all $k^2< 0$; this happens
because for $k^2<0$ one gets oscillatory solutions near the lower endpoint,
which for the $AdS$ case is at $r=0$.
Like confinement, the mass gap of the thermal $\N=4$ theory depends
on the cutoff at small $r$.  

\newsec{Approach To QCD}

One interesting way to study four-dimensional gauge theory is
by compactification from a certain exotic
six-dimensional theory with $(0,2)$ supersymmetry. This theory
can be realized in Type IIB compactification on K3
\ref\ewt{E. Witten, ``Some Comments On String Dynamics,'' in {\it Strings '95},
ed. I. Bars et. al. (World Scientific, 1997), hep-th/9507121.}
or in the dynamics of parallel $M$-theory fivebranes
\ref\oldstr{A. Strominger, ``Open $p$-Branes,'' Phys. Lett. {\bf B383} (1996)
44, hep-th/9512059.} and
can apparently be interpreted \malda\
in terms of $M$-theory on $AdS_7\times \S^4$.  This interpretation
is effective in the large $N$ limit -- as the $M$-theory radius of
curvature is of order $N^{1/3}$.  Since compactification from six to
four dimensions has been an effective approach to gauge theory dynamics
(for instance, in deducing Montonen-Olive duality \ewt\ using
a strategy proposed in \ref\oldduff{M. Duff, ``Strong/Weak Coupling
Duality From The Dual String,'' Nucl. Phys. {\bf B442} (1995) 47,
hep-th/9501030.}), it is natural
to think of using the solution for the large $N$ limit of the six-dimensional
theory as a starting point to understand the four-dimensional theory.

Our basic approach will be as follows.  If we compactify the six-dimensional
$(0,2)$ theory on a circle $C_1$ of radius $R_1$, with a
supersymmetry-preserving
spin structure (fermions are periodic in going around the circle), 
we get a theory that at low energies looks like
five-dimensional
$SU(N)$ supersymmetric Yang-Mills theory, with maximal supersymmetry
and five-dimensional gauge coupling constant $g^2_5=R_1$.  Now compactify
on a second circle $C_2$, orthogonal to the first, with radius $R_2$.  If
we take $R_2>>R_1$,  we can determine what the resulting four-dimensional
theory is in a two-step process, compactifying to five dimensions on
$C_1$ to get five-dimensional supersymmetric Yang-Mills theory and then
compactifying to four-dimensions on $C_2$.

No matter what 
spin structure we use on $C_2$, we will get a four-dimensional
$SU(N)$ gauge theory with gauge coupling $g_4^2=R_1/R_2$.  If
we take on $C_2$ (and more precisely, on $C_1\times C_2$)
the supersymmetry-preserving spin structure, then the
low energy theory will be the four-dimensional $\N=4$ theory some of
whose properties we have examined in the present paper.
We wish instead to break supersymmetry by taking the fermions to be
antiperiodic in going around $C_2$.  Then the fermions get masses
(of order $1/R_2$) at tree level, and the spin zero bosons very plausibly
get masses (of order $g_4^2N/R_2$) at one-loop level.  If this is so,
the low energy theory will be the pure $SU(N)$ theory without supersymmetry.
If $g_4^2<<1$, the theory will flow at very long distances to strong
coupling; at such long distances the spin one-half and spin one fields
that receive tree level or one-loop masses will be irrelevant.  So this
is a possible framework for studying the pure Yang-Mills theory
without supersymmetry.

We want to take the large $N$ limit with $g_4\to 0$ in such a way that
$\eta=g_4^2N$ has a limit.  So we need $g_4^2=\eta/N$, or in other words
\eqn\jurrt{R_1={\eta R_2\over N}.}
We actually want $\eta $ fixed
and small, so that the four-dimensional Yang-Mills theory is weakly coupled
at the compactification scale, and flows to strong coupling only at very
long distances at which the detailed six-dimensional setup is irrelevant.

To implement this approach, we first look for an Einstein manifold
that is asymptotic at infinity to $\R^5\times C_2$.  
Though it may seem to reverse the logic of the construction,
starting with $C_2$ first in constructing the solution turns out to be
more convenient.  The supersymmetry-breaking boundary
conditions on $C_2$ are the right ones for using the spacetime \oofball\
that is constructed by scaling of the seven-dimensional $AdS$ Schwarzschild
solution:
\eqn\ifball{ds^2=\left({\rho^2\over b^2}-{b^4\over\rho^{4}}\right)d\tau^2
+{d\rho^2\over \left({\rho^2\over b^2}-{b^4\over\rho^{4}}\right)} +
\rho^2\sum_{i=1}^{5}dx_i^2.}
According to \malda, we want here
\eqn\nifball{b=2G_N^{1/9}(\pi N)^{1/3}.}
(Here $G_N$ is the eleven-dimensional Newton constant, so $G_N^{1/9}$
has dimensions of length.  We henceforth set $G_N=1$.)   

To make the scaling with $N$ clearer,
we also set $\rho=2(\pi N)^{1/3}\lambda$.  And -- noting from \coofball\
that $\tau$ has period $4\pi b/n=(4/3)\pi^{4/3}N^{1/3}$ -- we set
\eqn\iccodo{\tau=\theta\cdot\left({2\pi N\over 3}\right)^{1/3},}
where $\theta$ is an ordinary angle, of period $2\pi$.  
After also a rescaling of the $x_i$, the metric becomes
\eqn\nicoco{ds^2={4\over 9}
\pi^{2/3}N^{2/3}\left(\lambda^2-{1\over \lambda^4}\right)
d\theta^2+4\pi^{2/3}N^{2/3}{d\lambda^2
\over \left(\lambda^2-
{1\over \lambda^4}\right)}+4\pi^{2/3}N^{2/3}\lambda^2\sum_{i=1}^5 dx_i^2.}

Now we want to compactify one of the $x_i$, say $x_5$, on a second circle
whose radius as measured at $\lambda=\infty$
should according to \jurrt\ should be $\eta /N$ times the radius of
the circle  parametrized by $\theta$.  To do this, we write $x_5=(\eta/N)\psi$
with $\psi$ of period $2\pi$.
We also now  restore the $\S^4$ factor that was present
in the original $M$-theory on $AdS_7\times \S^4$ and has so far been
suppressed.  The metric is now
\eqn\miccoc{\eqalign{ds^2=&
{4\over 9}
\pi^{2/3}N^{2/3}\left(\lambda^2-{1\over \lambda^4}\right)d\theta^2+
{4\over 9}\eta ^2\pi^{2/3}N^{-4/3}\lambda^2 d\psi^2\cr &+
4N^{2/3}{d\lambda^2
\over \left({\lambda^2}
-{1\over \lambda^4}\right)}+ 4\pi^{2/3}N^{2/3}\rho^2\sum_{i=1}^4 dx_i^2
+\pi^{2/3} N^{2/3} d\Omega_4^2.\cr}}
At this stage, $\theta$ and $\psi$ are both ordinary angular variables
of radius $2\pi$, and $d\Omega_4^2$ is the metric on a unit four-sphere.

Now, we want to try to take the limit as $N\to \infty$.  The metric
becomes large in all directions except that one circle factor -- the circle
$C_1$, parametrized by $\psi$ -- shrinks.  Thus we should try to use
the equivalence between $M$-theory compactified on a small circle
and weakly coupled Type IIA superstrings.  We see that the radius $R(\lambda)$
of the circle parametrized by $C$ is in fact
\eqn\coco{R(\lambda)={2\over 3}\eta \lambda \pi^{1/3}N^{-2/3}.}
To relate an $M$-theory  compactification on a circle to a Type IIA
compactification, we must \ref\uwitten{E. Witten, ``String Theory Dynamics
In Various Dimensions,'' Nucl. Phys. {\bf B443} (1995) 85, hep-th/9503124.} 
multiply the metric by $R$.  All factors
of $N$ felicitously disappear from the metric, which becomes
\eqn\miccoco{ds^2=
{8\over 27}\eta 
\lambda \pi\left({\lambda^2}-{1\over \lambda^4}\right)d\theta^2+
{8\pi\over 3} \eta \lambda {d\lambda^2
\over \left({\lambda^2}-
{1\over \lambda^4}\right)}+ {8\pi \over 3}\eta \lambda^3\sum_{i=1}^4 dx_i^2
+{2\pi\over 3} \eta \lambda d\Omega_4^2.}
The string coupling constant is meanwhile
\eqn\huccu{g_{st}^2=R^{3/2}={(2/3)^{3/2}
\eta ^{3/2}\lambda^{3/2}\pi^{1/2}\over N}.}

This result clearly has some of the suspected properties of large $N$
gauge
theories.  The metric \miccoco\ is independent of $N$, so in the weak
coupling limit, the spectrum of the string theory will be independent of
$N$.  Meanwhile, the string coupling constant \huccu\ is of order $1/N$,
as expected \thooft\ for the residual interactions between color singlet
states in the large $N$ limit.  
The very ability to get a description such as this one in which $1/N$ only
enters as a coupling constant (and not explicitly in the multiplicity of
states) is a reflection of confinement.  Confinement in the form of
an area law for Wilson loops can be demonstrated along the lines of
our discussion in section 3: it follows from the fact that the coefficient
in the metric of $\sum_{i=1}^4dx_i^2$ is bounded strictly above zero.
A mass gap likewise can be demonstrated, as in section 3, by using the
large $\lambda$ behavior of the metric.

On the other hand, it is not obvious how one could hope to compute the
spectrum or even show asymptotic freedom.
Asymptotic freedom should say that as $\eta \to 0$, the particle masses
become exponentially small (with an exponent determined by the gauge
theory beta function).  It is not at all clear how to demonstrate this.
A clue comes from the fact that the coupling of the physical hadrons
should be independent of $\eta $ (and of order $1/N$) as $\eta \to 0$. In view
of the formula \huccu, this means that we should take $\eta \lambda$
of order one as $\eta \to 0$.  If we set 
$\tilde \lambda=\eta \lambda$, and write
the metric in terms of $\tilde \lambda$, then the small $\eta $ limit becomes
somewhat clearer: a singularity develops at small $\tilde\lambda$ for
$\eta \to 0$.  Apparently, in this approach, the mysteries of four-dimensional
quantum gauge theory are encoded in the behavior of string theory near
this singularity.   

This singularity actually has a very simple and intuitive interpretation
which makes it clearer why four-dimensional gauge theory can be described
by string theory in the spacetime \miccoco.  The Euclidean signature
Type IIA nonextremal
fourbrane solution is described by the metric \ref\horstro{G. Horowitz
and A. Strominger,  ``Black Strings And $p$-Branes,'' Nucl. Phys.
{\bf B360} (1991) 197.}
\eqn\ucx{\eqalign{ds^2= & \left(1-\left({r_+\over r}\right)^3\right)
\left(1-\left({r_-\over r}\right)^3\right)^{-1/2}dt^2
+\left(1-\left({r_+\over r}\right)^3\right)^{-1}
\left(1-\left({r_-\over r}\right)^3\right)^{-5/6}{dr^2}\cr &
+\left(1-\left({r_-\over r}\right)^3\right)^{1/2}\sum_{i=1}^4 dx_i^2
+r^2\left(1-\left({r_-\over r}\right)^3\right)^{1/6}d\Omega_4^2,\cr}}
with $r_+>r_->0$.
The string coupling constant is
\eqn\hubbo{g_{st}^2= \left(1-\left({r_-\over r}\right)^3\right)^{1/2}.}
The horizon is at $r=r_+$, and the spacetime is bounded by $r\geq r_+$. 
This spacetime is complete and smooth if $t$ has period 
\eqn\uzz{T=12\pi\left(1-\left({r_-\over r_+}\right)^3\right)
^{-1/6}.}
 If one continues (via Lorentzian or complex values
of the coordinates) past $r=r_+$, there is a singularity at $r=r_-$.
The extremal fourbrane solution is obtained by setting $r_+=r_-$
and is singular.  But this singularity is exactly the singularity
that arises in \miccoco\ upon taking $\eta\to 0$, with $\eta\lambda\sim 1$!
In fact, if we set $\lambda^6=(r^3-r_-^3)/(r_+^3-r_-^3)$, identify
$\eta$ with $(1-(r_-/r_+)^3)^{1/6}$, and take
the limit of $r_+\to r_-$, then \ucx\ reduces to \miccoco, up to some
obvious rescaling.
  Moreover, according to \uzz, $r_+\to r_-$ is the limit
that $T$ is large, which (as $1/g_4^2=T/g_5^2$) makes the four-dimensional
coupling small.

So in hindsight we could discuss four-dimensional gauge theories in the
following way, without passing through the CFT/$AdS$ correspondence.
In a spacetime $\R^9\times \S^1$, consider $N$ Type IIA fourbranes wrapped
on $\R^4\times \S^1$.  Pick a spin structure on the $\S^1$ that breaks
supersymmetry.  This system looks at low energies like four-dimensional
$U(N)$ gauge theory, with Yang-Mills coupling $g_{4}^2=g_5^2/T=g_{st}/T$.  Take
$N\to\infty$ with $g_{st}N$ fixed.   The $D$-brane system has both open
and closed strings.  The dominant string diagrams for large $N$ with
fixed $g_5^2N$ and fixed $T$
 are the planar diagrams of 't Hooft \thooft\ -- diagrams
of genus zero with any number of holes.  (This fact was exploited recently
\nek\ in analyzing the beta function of certain field theories.)  Summing
them up is precisely the long-intractable problem of the $1/N$ expansion.

Now, at least if $\eta$ is large, supergravity effectively describes
the sum of planar diagrams in terms of the metric \ucx\ which is produced
by the $D$-branes.
This is a smooth metric, with no singularity and no $D$-branes.
So we get a description with closed Type IIA strings only.  Thus
the old  prophecy \thooft\ is borne out: nonperturbative effects
 close up the holes in the Feynman diagrams, giving a confining
theory with a mass gap, and with $1/N$ as a coupling constant,
at least for large $\eta$.  To understand large
$N$ gauge theories, one would want, from this point of view, to
show that there is no singularity as a function of $\eta$, except at $\eta=0$,
and to exhibit asymptotic freedom and compute the masses
for small $\eta$.  (This looks like a tall order, given our limited
knowledge of worldsheet field theory with Ramond-Ramond fields in the
Lagrangian.)  The singularity
at $\eta=0$ is simply the singularity of the fourbrane metric at $r_+=r_-$;
it reflects the classical $U(N)$
gauge symmetry of $N$ parallel fourbranes, which disappears quantum
mechanically when $\eta\not= 0$ and the singularity is smoothed out.

\bigskip
I have benefited from comments by N. Seiberg.
This work was supported in part by NSF Grant PHY-9513835.
\listrefs
\end